# Topic Modelling of Everyday Sexism Project Entries

Sophie Melville[1], Kathryn Eccles[1], and Taha Yasseri[1,2]*

[1]Oxford Internet Institute, University of Oxford, Oxford, UK

[2]Alan Turing Institute, London, UK

**\*Correspondence:**
Taha Yasseri
taha.yasseri@oii.ox.ac.uk

**Abstract**
The Everyday Sexism Project documents everyday examples of sexism reported by volunteer contributors from all around the world. It collected 100,000 entries in 13+ languages within the first 3 years of its existence. The content of reports in various languages submitted to Everyday Sexism is a valuable source of crowdsourced information with great potential for feminist and gender studies. In this paper, we take a computational approach to analyze the content of reports. We use topic-modelling techniques to extract emerging topics and concepts from the reports, and to map the semantic relations between those topics. The resulting picture closely resembles and adds to that arrived at through qualitative analysis, showing that this form of topic modeling could be useful for sifting through datasets that had not previously been subject to any analysis. More precisely, we come up with a map of topics for two different resolutions of our topic model and discuss the connection between the identified topics. In the low resolution picture, for instance, we found Public space/Street, Online, Work related/Office, Transport, School, Media harassment, and Domestic abuse. Among these, the strongest connection is between Public space/Street harassment and Domestic abuse and sexism in personal relationships.The strength of the relationships between topics illustrates the fluid and ubiquitous nature of sexism, with no single experience being unrelated to another.

## 1   Introduction

"Women across the country - and all over the world, in fact - are discovering new ways to leverage the internet to make fundamental progress in the unfinished revolution of feminism" - #FemFutureReport (Femfuture, 2017).
Laura Bates, founder of the Everyday Sexism Project, has signalled that "it seems to be increasingly difficult to talk about sexism, equality, and women's rights" (Bates, 2015). With many theorists suggesting that we have entered a so-called 'post-feminist' era in which gender equality has been achieved (McRobbie, 2009), to complain about sexism not only risks being labelled as "uptight", "prudish", or a "militant feminist", but also exposes those who speak out to sustained, and at times vicious, personal attacks (Bates, 2015). Despite these risks, thousands of women are speaking out about their experiences of sexism, and are using digital technologies to do so (Martin & Valenti, 2012), leading to the development of a so-called 'fourth wave' of feminism, incorporating a range of feminist practices that are enabled by Web 2.0 digital technologies (Munro, 2013).
The 'Everyday Sexism Project', founded by Bates in 2012, is just one of the digital platforms employed in the fight back against sexism. Since its inception, the site has received over 100,000 submissions in more than 13 different languages, detailing a wide variety of experiences. Submissions are uploaded directly to the website, and via the Twitter account @EverydaySexism and hashtag #everydaysexism. Until now, analysis of posts has been largely qualitative in nature, and there has been no systematic analysis of the nature and type of topics discussed, or whether distinct 'types' of sexism emerge from the data. In this paper, we expand the methods used to investigate Everyday Sexism submission data by undertaking a large-scale computational study, with the aim of enriching existing qualitative work in this area (Becker and Swim, 2011; Swim et al. 2001). To the best of our knowledge this is the first time a dataset at this scale is being analysed to come up with a data-driven typology of sexism. It is important to note, however, that the data under study suffer from intrinsic biases of self-reported experiences that might not represent a complete picture of sexism..

Our analysis of the data is based on Natural Language Processing, using topic-modelling techniques to extract the most distinctly occurring topics and concepts from the submissions. We explored data-driven approaches to community-contributed content as a framework for future studies. Our research seeks to draw on the rich history of gender studies in the social sciences, coupling it with emerging computational methods for topic modelling, to better understand the content of reports to the Everyday Sexism Project and the lived experiences of those who post them.

The analysis of "prejudice, stereotyping, or discrimination, typically against women, on the basis of sex" (Oxford English Dictionary, 2016) has formed a central tenet of both academic inquiry and a radical politics of female emancipation for several decades[1]. Studies of sexism have considered it to be both attitudinal and behavioural, encompassing both the endorsement of oppressive beliefs based on traditional gender-role ideology, and what we might term more 'formal' discrimination against women on the basis of their sex, for example in the workplace or in education (Harper, 2008, p. 21). Peter Glick and Susan T. Fiske build on this definition of sexism in their seminal 1996 study *The Ambivalent Sexism Inventory: Differentiating Hostile and Benevolent Sexism*, where they present a multidimensional theory of sexism that encompasses two components: 'hostile' and 'benevolent' sexism. As the authors highlight, traditional definitions of sexism have conceptualised it primarily as a reflection of hostility towards women, but this view neglects a significant further aspect of sexism: the "subjectively positive feelings towards women" that often go hand in hand with sexist apathy (Glick & Fiske, 1996, p. 493).

More recent studies, particularly in the field of psychology, have shifted the focus away from *who* experiences sexism and *how* it can be defined, towards an examination of the psychological, personal, and social implications that sexist incidents have for women. As such, research by Buchanan & West (2009), Harper (2008), Moradi & Subich (2002), and Swim et al (2001) has highlighted the damaging intellectual and mental health outcomes for women who are subject to continual experiences of sexism. Moradi and Subich, for example, argue that sexism combines with other life stressors to create significant psychological distress in women, resulting in low self-esteem and the need to "seek therapy, most commonly for depression and anxiety" (Moradi & Subich, 2002, p. 173). Other research indicates that a relationship exists between experiences of sexism over a woman's lifetime and the extent of conflict she perceives in her romantic heterosexual relationships (Harper, 2008); that continual experiences of sexism in an academic environment results in women believing that they are inferior to men (Ossana, Helms, & Leonard, 1992); and that disordered eating among college women is related to experiences of sexual objectification (Sabik and Tylka, 2006).

Given its increasing ubiquity in everyday life, it is hardly surprising that the relationship between technology and sexism has also sparked interest from contemporary researchers in the field. Indeed, several studies have explored the intersection between gender and power online, with Susan Herring's work on gender differences in computer-mediated communication being of particular note (cf. Herring, 2008). Feminist academics have argued that the way that women fight back against sexism in the digital era is fundamentally shaped by the properties, affordances, and dynamics of the 'web 2.0' environments in which much current feminist activism takes place, with social media sites uniquely

---

[1] Cf. De Beauvoir, 1949; Friedan, 1963; Firestone, 1970; Hartstock, 1983; hooks, 1984

facilitating "communication, information sharing, collaboration, community building and networking" in ways that neither the static websites of Web 1.0 nor the face-to-face interactions of earlier feminist waves have been able to (Jessalyn Keller, 2012; Carstensen, 2009).

Theorists in the field of psychology have focused on the impact that using digital technology, and particularly Web 2.0 technologies, to talk about sexism can have on women's well-being. Mindi D. Foster's 2015 study, for example, found that when women tweeted about sexism, and in particular when they used tweets to a) name the problem, b) criticise it, or c) to suggest change, they viewed their actions as effective and had enhanced life satisfaction, and therefore felt empowered (Foster, 2015: 21). These findings are particularly relevant to this study, given the range of channels offered by the Everyday Sexism project to those seeking to 'call out' sexism that they've experienced or witnessed both online and off.

Despite the diversity of research on sexism and its impact, there remain some notable gaps in understanding. In particular, as this study hopes to highlight, little previous research on sexism has considered the different and overlapping ways in which sexism is experienced by women, or the sites in which these experiences occur, beyond an identification of the workplace and the education system as contexts in which sexism often manifests (as per Barnett, 2005; Watkins et al., 2006; Klein, 1992). Furthermore, research focusing on sexism has thus far been largely qualitative in nature. Although a small number of studies have employed quantitative methods (cf. Brandt 2011; Becker and Wright, 2011), none have used computational approaches to analyse the wealth of available online data on sexism. Here we seek to fill such a gap. By providing much needed analysis of a large-scale crowd sourced data set on sexism, it is our hope that knowledge gained from this study will advance both the sociological understanding of women's lived experiences of sexism, and methodological understandings of the suitability of computational topic modelling for conducting this kind of research. In other research topic modelling has been extensively used (Puschmann & Scheffler 2016) to study the history of computational linguistics (Hall, Jurafsky and Manning, 2008), U.S. news media in the wake of terror attacks (Bonilla and Grimmer, 2013), online health discourse (Ghosh and Guha, 2013; Paul and Dredze, 2014), historical shifts in news writing (Yang, Torget & Mihalcea ,2011), political discourse (Koltsova and Koltcov, 2013), and online electoral campaigns (McElwee and Yasseri 2017). In this project we are interested in discussing in particular, what the emerging topics can tell us about the ways in which sexism is manifested in everyday life.

## 2   Data and Methods

We collected the content of posts on the Everyday Sexism website in February 2015, with the permission of the website owner, through a simple web crawler. The project adhered at all times to the Oxford Central University Research Ethics Committee's (CUREC) Best Practice Guidance 06_Version 4.0 on Internet-Based Research (IBR). None of the project entries is quoted in the article, and the approach deliberately looked for patterns and connections rather than isolating individual accounts or contributors.

In processing the data, after cleaning the posts that were not in English, we ended up with 78,783 posts containing 3,221,784 words. We then removed all punctuation and English language stop-words (such as 'and', 'it', 'in' etc) from the data, this is a standard practice in the literature (Wallach, Mimno & McCallum, 2009). The data were then split into individual words, which were stemmed using an nltk English language snowball stemmer (Perkins, 2010).

Topic modelling is a technique that seeks to automatically discover the topics contained within a group of documents. 'Documents' in this context could refer to text items as lengthy as individual books, or as short as sentences within a paragraph. For instance, if we assumed that each sentence of a corpus of text is a "document" we would have:

- Document 1: I like to eat kippers for breakfast.
- Document 2: I love all animals, but kittens are the cutest.
- Document 3: My kitten eats kippers too.

We therefore assume that each sentence contains a mixture of different topics and that a 'topic' is a collection of words that are more likely to appear together in a document.

The algorithm is initiated by setting the number of topics that it needs to extract. It is hard to guess this number without having insight into the topics, but one can think of this as a resolution tuning parameter. The smaller the number of topics is set, the more general the bag of words in each topic would be, and the looser the connections between them.

The algorithm loops through all of the words in each document, assigning every word to one of the topics in a temporary and semi-random manner. This initial assignment is arbitrary and it is easy to show that different initializations lead to the same results in long run. Once each word has been assigned a temporary topic, the algorithm then re-iterates through each word in each document to update the topic assignment using two criteria: 1) How prevalent is the word in question across topics? and 2) How prevalent are the topics in the document?

To quantify these two, the algorithm calculates the likelihood of the words appearing in each document assuming the assignment of words to topics (word-topic matrix) and topics to documents (topic-document matrix). Words can appear in different topics and more than one topic can appear in a document. But the iterative algorithm seeks to maximize the self-consistency of the assignment by maximizing the likelihood of the observed word-document statistics.

We can illustrate this process and its outcome by going back to the example above. A topic modelling approach might use the process above to discover the following topics across our documents:

- Document 1: I like to eat kippers for breakfast. [100% Topic A]
- Document 2: I love all *animals*, but *kittens* are the cutest. [100% Topic B]
- Document 3: My *kitten* eats kippers too. [67% Topic A, 33% Topic B],

where

- Topic A: eat, kippers, breakfast
- Topic B: *animals, kittens*

Topic modelling defines each topic as a so-called 'bag of words', but it is the researcher's responsibility to decide upon an appropriate label for each topic based on their understanding of language and context. Going back to our example, the algorithm might classify the underlined words under Topic A, which we could then label as 'food' based on our understanding of what the words mean. Similarly, the *italicised* words might be classified under a separate topic, Topic B, which we could label 'animals'. In this simple example the word "eat" has appeared in a sentence dominated by Topic A, but also in a sentence with some association to Topic B. It can therefore be seen as a connector of the two topics.

We used a similar approach to first extract the main topics reflected in the reports made to the Everyday Sexism Project website, and then extracted the relation between the sexism-related topics and concepts based on the overlap between the bags of words of each topic. For this we used a simple implementation of the LDA algorithm for topic modelling (Pritchard, Stephens, & Donnelly, 2000 and Blei, Ng, & Jordan, 2003).

## 3 Results

Tables 1 and 2 list the topics that are detected by topic modelling algorithm for two different numbers of topics n = 7 and n = 20. Each row shows the top 50 words that are most prevalent in each topic. The 3rd column shows the qualitative annotations. By increasing the number of topics, we will have less granularity however, the annotation task becomes more difficult as topics become more diverse. However, combining the two pictures we shed light on the most apparent images of sexism as reported on the Everyday Sexism website.

**Table 1.**
**The topics extracted from the Everyday Sexism website computationally and annotated qualitatively for n=7.**

| Topic Number | Assigned Words | Annotation |
| --- | --- | --- |
| S0 | friend man guy one hand away back tri get walk look said grab time start got go around felt ask told stop say us bus next like behind night happen went turn could touch would came sit feel even move way put veri bar train know want club tell face | Public space/ Street harassment |
| S1 | women men becaus like make woman feel think peopl sexism get say thing comment male would even know one man want sexual way femal whi mani veri time sexist onli realli ani friend also girl never look much someth made person joke need call tell right seem use tri life | Online/ Comments |
| S2 | work male ask job one said femal told colleagu manag would get husband boss man time woman becaus offic onli women compani name say look need year go call custom day want meet new make even could take men staff got help first know pay question talk boyfriend use marri | Work/office /company/ customer |
| S3 | walk car man men street shout home guy get look one call past go stop us time friend said got follow way drive yell road day back like say two ask whistl start pass around driver bus group window run work wear old make turn park even fuck feel bike | Transport/ Street harassment |
| S4 | boy girl school wear year told becaus class like one teacher said look old would get dress ask male friend day say make onli guy play hair student go short want age high skirt even time thing us got comment femal shirt group whi tell realli call good laugh cloth | School/ Teacher/ Uniform |
| S5 | women girl men woman femal like look show whi onli male man play one game say watch pictur read comment get ad articl pink see new facebook love photo page boy buy advert today shop post video news book use mum footbal http sexist sport magazin tv becaus everydaysex ladi | Media |
| S6 | friend told becaus want would said year time go guy ask one get like boyfriend tell say rape never know tri even got sex happen start went hous girl home feel still thing day talk could thought realli make night call brother think sexual stop onli back old dad made | Domestic abuse/ Relationships/Home |

Figure 1 shows the number of posts that are primarily assigned to each topic for n = 7 and 20 respectively. One should note that, because of the way in which topic modelling was implemented in this work, topics would emerge with comparable sizes in terms of the number of documents assigned to them. Hence these histograms might be biased and considering the fact that the original dataset has its own natural biases of self-reported sample, the frequency analysis cannot be used to draw any conclusions.

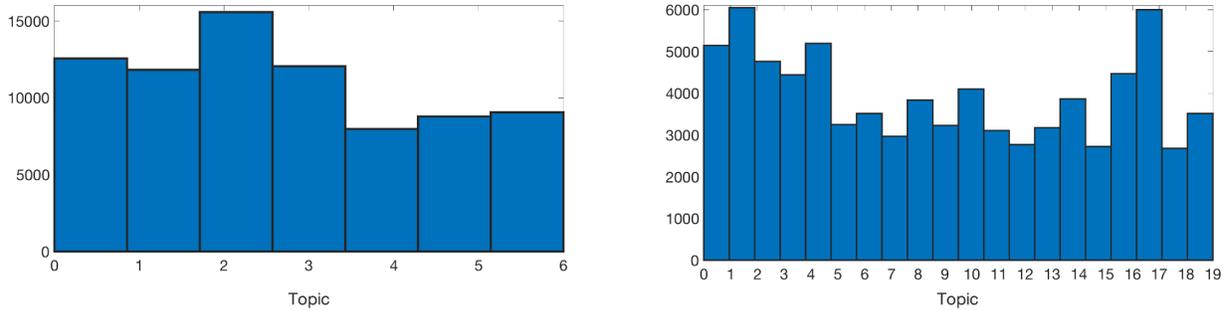

**Figure 1.**
**The number of posts that are primarily assigned to each topic for n =7 (left) and 20 (right).**

In the next step, we consider the similarity between topics. This can be done in two ways: 1 by comparing how words are assigned to each pairs of topics and 2) by comparing how documents are assigned to each topic. The first approach is more suitable when we have smaller number of topics and hence larger overlap between the words assigned to each topic whereas the second approach can be used when there are more topics and each document is assigned to multiple topics at the same time.
We quantified these similarities by calculating the *cosine similarity* between the vectors of word weights and topic weights in the word-topic and topic-document matrixes. Then we used the cosine similarity as the weight of the connection between topics as depicted in Figure 2 for n = 7 and 20.

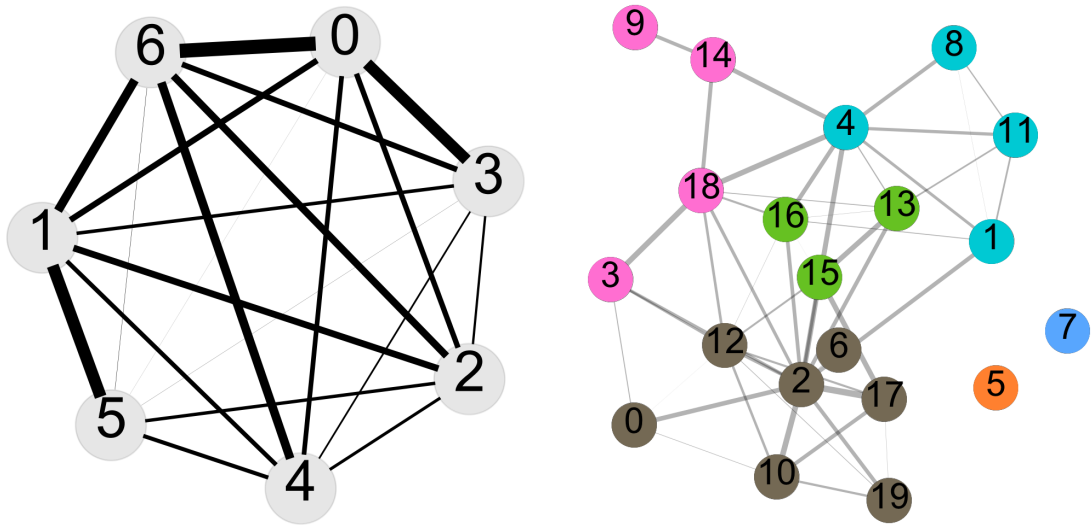

**Figure 2.**
**Network visualizations of the topics for n =7 (left) and 20 (right). The weight of the connections between pairs of the topics is based on the similarity of how the words are assigned to them in the left panel and how topics are assigned in the right panel. We removed connections with cosine similarities smaller than 0.2 for clarity. The colour-code of the right panel is based on the communities that are detected using the Gephi implementation of the Leuven algorithm of community detection in networks.**

In the case of 20 topics, we can also try to cluster topics into groups based on simple clustering algorithms in network science that group nodes of a network based on the strengths of their connections. The right panel of Figure 2 shows such grouping based on the Leuvain algorithm (Blondel, Guillaume, Lambiotte, & Lefebvre, 2008) as implemented in *Gephi* (Bastian, Heymann, & Jacomy, 2009) with the resolution 1.0 in the settings.

**Table 2**
**The topics extracted from the Everyday Sexism website computationally and annotated qualitatively for n=20.**

| Topic Number | Assigned Words | Annotation |
|---|---|---|
| L0 | friend guy grab one night club hand away tri walk danc bar around us back man turn get told group grope go touch behind time went laugh start put came | Socialising |
| L1 | work male job colleagu manag boss femal offic one ask told compani would meet onli staff said worker day year time woman becaus women new interview get team senior | Work |
| L2 | feel like make think would becaus say even know thing time realli someth made peopl felt veri look want way comment never happen one thought get go tri could uncomfort | Comments |
| L3 | friend told becaus want guy boyfriend would one said sex time go ask tri night get got tell start like went even know never say thought year room sleep back | Domestic abuse |
| L4 | women men sexism woman becaus peopl like male make femal think sexist man way even mani comment thing gender feminist get feel equal whi onli also societi seem say veri | Feminism |
| L5 | work help man need get said ask woman car one becaus drive use look know told say women weight could men go put would lift like clean thing guy onli | Other |
| L6 | work ask man custom shop said male drink bar one boyfriend look restaur store get order pay buy went time told tabl men food hand say eat serv like card | Customer/ Workplace |
| L7 | name husband ask doctor call male address nurs phone first onli partner said whi car new chang get question mrs mr email account hous even marri went use boyfriend miss | Titles, forms of address |
| L8 | husband get want children becaus mother marri work told famili ask dad father woman man women home kid go job mum wife time would babi cook whi tell parent need | Workplace/ Parenting/ Home |
| L9 | girl play boy game like footbal one pink team femal sport watch onli toy male becaus littl women video daughter love want charact gender show whi book music player band | Sport / Media |
| L10 | walk home man go ask follow back away friend us start get one around said street alon car guy stop tri look call could got door leav time night way | Street harassment |
| L11 | male student femal univers one women class ask studi said becaus work told onli year colleg girl talk would group cours scienc question well first engin time lectur good woman | University |
| L12 | year old told age sister time brother older dad said would man friend like never girl one famili boy mother tell look parent mom went still us start day happen | Home/ Families |
| L13 | boy school girl teacher year class one told would friend becaus old said high us day call age group onli like ask laugh grade male even thing make time student | School |
| L14 | women men pictur facebook comment articl post read show page photo femal news http look magazin ad nake uk whi today watch see www com websit everydaysex male advert | Social media/ Media |
| L15 | wear dress look short like hair becaus skirt cloth shirt men make top get told comment day leg bodi feel breast girl wore jean want even whi long cut show | Clothing /Appearance |
| L16 | guy say said like friend girl told get call becaus ask joke want one talk know look tell fuck whi think man make bitch realli woman thing got laugh | Street Harassment |
| L17 | walk car shout men street past man home whistl guy yell road call drive group window pass get one stop two van bike way us got day run friend driver | Street Harassment |
| L18 | rape sexual harass abus men assault becaus women polic happen would year time get peopl report mani feel victim man woman know stori told even tri ani live never experi | Assault/Violence |
| L19 | man bus train next look hand sit stop got move sat back away felt get seat start one tri leg behind could stand said stare touch around time turn wait | Public Transport |

## 4   Discussion and Conclusions

Analysis of the Everyday Sexism data has hitherto largely been qualitative in nature, with themes and sites associated with experiences of sexism drawn out in Bates' book, *Everyday Sexism* (Bates, 2014) and journalism. In her book, Bates identifies common sites of sexism drawn from the Everyday Sexism submissions, which include: Young Women Learning, Women in Public Spaces, Women in the Media, Women in the Workplace, and Motherhood (which we might also read as Women in the Home).[2] More recently the Everyday Sexism website introduced a new option for tagging experiences using the following groupings: Workplace, Public Space, Home, Public Transport, School, University, Media. In the topics that emerge from our analysis of the Everyday Sexism accounts, these same areas are essentially replicated, referring in the first analysis of seven topics (Table 1) to Young Women Learning (Topic S4), Women in Public Spaces (Topics S0 and S3), Women in the Media (Topics S1 and S5), Women in the Workplace (Topic S2) and Women in the Home (Topic S6). This finding bears out the qualitative categorisations of the data set by Bates, and offers an important understanding of how topic modelling could be useful in processing and beginning to understand similar data sets that have not yet been analysed.

One area that does not appear as a discrete category in Bates' book, or in the tags on the Everyday Sexism website, is something that we have categorised in n = 7 as 'online' sexism, or 'comments'. It appears in our analysis as a separate topic, S1, with the word 'onli' also appearing in four other topics:

---

[2] Bates is careful to include outlying, or less common yet equally relevant and important experiences of sexism. We have taken care to avoid a quantitative approach that might count and rank experiences of sexism from most common (and therefore important) to least.

school, work, media and home. Although the Everyday Sexism accounts are submitted through the website, or via Twitter, the purpose of the site is to log everyday instances of sexism, both on and offline, and the majority of topics relate to offline experiences of sexism. One of the main findings from this study is that experiences of sexism, even loosely grouped in the ways that we have described, are located everywhere. They are connected and they are pervasive. The appearance of 'online' as a separate topic, together with its appearance in four of the seven n = 7 topics, suggests the prevalence of sexism as mediated through digital tools, with the 'online' sphere constituting a quasi-public, quasi-private 'space' in which sexism can be enacted, and as a nexus through which sexist abuse enacted in other spheres can be continued and reinforced.

When we increase the number of topics to 20 (Table 2), this allows us to break down these experiences into separate but connected sites of sexism. For example, young women are clearly experiencing sexism in their learning environments, as evident in Topic S4 of our initial analysis but in the larger sample, we see sexism being experienced in both the school and University (Topics L11 and L13), areas connected by being associated with learning and with formative experiences of gender relations and expectations. The patterns of sexism experienced in the classroom at school may well pave the way for similar behaviour in the lecture hall or university classroom, with the majority of words in both topics overlapping. We also see issues around gender and sport surfacing (Topic L9), with 'girl', 'boy', 'football', 'sport' and 'pink' suggesting gendered notions of what constitutes appropriate forms of exercise and recreation, and reflecting the early age at which these gender stereotypes are operational (Eccles et al, 1990). Subtle differences in the ways in which these educational, professional and leisure spaces operate can be exposed by this more finely tuned analysis.

In our analysis of the larger number of topics (n=20), work becomes a more complex setting for types of sexism, with topics L1, L5 and L8 all referring to the workplace as a site of sexism, either through 'manager' 'boss' or 'colleague', or through the division of domestic labour in the home, where we see 'job' and 'work' being juxtaposed with 'mother' and 'father', 'children' and 'kid', and 'husband' and 'wife'. In this way, we see the layering of experiences of sexism in the public sphere of work, education and business on top of sexism experienced at home, with inequalities in the workforce perhaps compounded by inequalities in the division of household and parenting tasks. The home is a hugely influential space in which children begin to witness and absorb expectations around gendered roles and behaviour. Bates often refers to this as a type of 'institutional sexism', and argues that these early experiences can shape and dictate a woman's interests, activities and behaviour. (Bates, 2014). Topic L12 draws together a picture of sexism in the family, and hints at the beginnings of victim blaming in the clustering of these familial roles ('brother', 'dad', 'sister', 'mother') with 'would', 'start' and 'happen'. A perhaps subtler form of sexism, still within the home, is expressed in topic L7, where titles and forms of address are prevalent, reflecting the ways in which these can become 'vehicles by which people establish or contest their positions within communities of practice' (Mills, 2003).

Analysis of the larger number of topics draws out numerous topics associated with what we may cluster together as street harassment, or Women in Public Spaces. Separating these topics out allows us to arrive at a more complex view of the reports that generate these clusters. Topics 10 and 17 suggest the frequency of accounts of women being verbally harassed, followed and threatened in the street while simply going about their daily lives. Topic 16 is suggestive of the frequency with which women are expected to 'laugh' this off as a 'joke', and the co-location of 'bitch' and 'fuck' suggests that women's reluctance to do so is also not met with levity. This is a theme drawn out by Bates' qualitative reading of the accounts, and evokes the close relationship between what Glick and Fiske refer to as 'benevolent sexism' and 'hostile sexism', and the way in which the former (seeking positive reinforcement) quickly becomes the latter, further reinforcing the connectivity between these different accounts (Bates, 2014; Glick and Fiske, 1996). Topic L19, which clusters 'bus', 'train', 'stop' and 'seat' suggests that using public transport offers no defence against experiences of everyday sexism, with 'hand', 'felt', 'leg', 'behind', 'stare' and 'touch' indicative of experiences commonly identified by victims of sexual assault. Topic L18, in which we find 'rape', 'sexual harass', 'abuse', 'assault', 'police' and 'victim' creates a stark picture of the culmination of these threatening behaviours.

It is also possible to extract themes in the data through relationships between topics exposed through our analysis, shown in Figure 2. In the smaller group of topics (n=7), the relationship strength is shown through the thickness of the connective lines. Topics S0 (Public space/Street harassment) and S3 (Transport/Street harassment) have a strong and obvious connection, as do topics S1 (Online/Comments) and S5 (Media). Other connections are superficially less clear. Topics S0 (Public space/Street harassment) and S6 (Home/Relationships), for example, are very strongly connected. While this may seem baffling at first glance, it ties in with Bates' observations of how sexism is reinforced in the home when victims of sexual harassment are subject to judgment and blame when reporting incidents to those close to them (Bates, 2014, p. 34-41).

For the larger group of topics (n=20), relationships are depicted through the different coloured groupings (Figure 2). The groups are identified based on the strength of the connections between topics assessed through the overlap of documents co-assigned to them. This picture shows how various sub-topics are interconnected and the experience of sexism is not isolated in one shape or form. However, the two topics, L5 and L7, appear unconnected to the other topics, with no ties either strong or weak. In fact, these topics appear to be quite general, remaining both distanced from and yet relevant to other topics. In Topic L5, it is difficult to categorise this group of words into a discrete topic, as signifying words such as 'work', 'drive', 'clean', 'weight', 'car' are difficult to cluster. In Topic L7, the use or misuse of appropriate titles and forms of address are experienced as everyday sexism, which may emerge across a range of backdrops. The presence of topics L5 and L7, alongside topic S1 (online/comments) in our n=7 sample, serves to remind us that sexism can be both focused, on particular sites, roles and activities, and all encompassing, bypassing neat categorisation.

What can topic modelling of the Everyday Sexism data set tell us about experiences of sexism? The topic modelling approach delivers word bags containing highly distilled elements of commonly experienced sexist encounters, creating stark pictures of interrelated sites, languages and relationships in which sexism is enacted. This analysis suggests that sexism is fluid; it's not limited to a certain space, class, culture, or time. It takes different forms and shapes but these are connected. Sexism penetrates all aspects of our lives, it can be subtle and small, and it can be violent and traumatising, but it is rarely an isolated experience.

What does this method add to a qualitative analysis, and how can this sort of study be useful? In summary, topic modelling provides an effective means of analysing a large data set to produce high level as well as subtler and more finely drawn themes and commonalities. Using a data set like the Everyday Sexism reports, which have already been subject to extensive qualitative analysis, allows us to test this method against qualitative findings, producing consistent results. One concern at the beginning of this project was that this method may seem reductive, producing the most common and therefore, it could be argued, most affecting or important experiences or sites of sexism. The topic modelling approach in fact offers a largely inclusive set of findings, highlighting distinct topics but visualising connections between these topics, providing the opportunity to tease out connected but subtly different topics, which can then be contextualised by qualitative readings of the reports.

The results presented here are based on preliminary analysis, but in the future a more sophisticated approach both in the sense of methods of topic modelling and using larger and more representative datasets could potentially improve the results significantly. This could allow researchers to use computational methods to extract concepts and patterns that could then inform policy agendas.

## 5  Conflict of Interest

*The authors declare that the research was conducted in the absence of any commercial or financial relationships that could be construed as a potential conflict of interest.*

## 6  Author Contributions

KE and TY designed the research. SM and TY performed the computational analysis. KE performed the qualitative analysis. SM, KE, and TY wrote the manuscript.

## 7  Funding

This publication arises from research funded by the John Fell Oxford University Press (OUP) Research Fund. TY was partially supported by The Alan Turing Institute under the EPSRC grant EP/N510129/1.

## 8  Acknowledgments

We would like to thank Laura Bates for useful comments and suggestions throughout the project.